\documentclass[pdflatex,sn-basic]{sn-jnl}

\jyear{2021}%

\theoremstyle{thmstyleone}%
\theoremstyle{thmstyletwo}%

\theoremstyle{thmstylethree}%

\usepackage{etoolbox}   
\makeatletter
\patchcmd{\@maketitle}{\artauthors}{{\artauthors}}{}{}
\makeatother
\begin{document}

\title[Withdrawal Success Optimization in a Pooled Annuity Fund]{Withdrawal Success Optimization in a Pooled Annuity Fund}


\author[1]{\fnm{\quad\quad\quad\quad Hayden} \sur{Brown} ORCID: 0000-0002-2975-2711}\email{haydenb@nevada.unr.edu}

\affil[1]{\orgdiv{Department of Mathematics and Statistics}, \orgname{University of Nevada, Reno}, \orgaddress{1664 \street{N. Virginia Street}, \city{Reno}, \postcode{89557}, \state{Nevada}, \country{USA}}}


\abstract{Consider a closed pooled annuity fund investing in $n$ assets with discrete-time rebalancing. At time 0, each annuitant makes an initial contribution to the fund, committing to a predetermined schedule of withdrawals. Require annuitants to be homogeneous in the sense that their initial contributions and predetermined withdrawal schedules are identical, and their mortality distributions are identical and independent. Under the forementioned setup, the probability for a particular annuitant to complete the prescribed withdrawals until death is maximized over progressively measurable portfolio weight functions. Applications consider fund portfolios that mix two assets: the S\&P Composite Index and an inflation-protected bond. The maximum probability is computed for annually rebalanced schedules consisting of an initial investment and then equal annual withdrawals until death. A considerable increase in the maximum probability is achieved by increasing the number of annuitants initially in the pool. For example, when the per-annuitant initial contribution and annual withdrawal amount are held constant, starting with 20 annuitants instead of just 1 can increase the maximum probability (measured on a scale from 0 to 1) by as much as .15.}







\maketitle

\section{Introduction}\label{sec1}
After a lump sum investment, suppose an investor wishes to complete a pre-determined schedule of withdrawals. In practice, the total amount intended to be withdrawn is usually larger than the lump sum. To overcome this disparity, the lump sum must be invested in various assets having positive expected log-returns, with the goal of maximizing the probability to complete the schedule of withdrawals. Note that this maximization occurs over the time-adapted portfolio weight functions, which control the proportion of available wealth invested in each asset. For the case where rebalancing and withdrawals are made discretely in time, this maximization problem is studied for an individual investor in \cite{brown}. The goal here is to extend the results of \cite{brown} to a pool of investors, all hoping to make the same schedule of withdrawals. Of particular interest is the schedule having a constant annual withdrawal until death.

The basic idea of the pooled annuity fund is as follows. At time 0, collect the same lump sum investment from each individual in the pool. Invest the combined funds into a fixed number of assets, rebalancing and withdrawing periodically. When a member of the pool dies, no funds are removed from the pool for a beneficiary. Instead, the dead member's share of the combined funds is kept in the fund to benefit the remaining living members of the pool. In effect, the members of the pool are mutual beneficiaries, coming together to insure against longevity risk. 

These pooled annuity funds can also be called tontines. The word \textit{tontine} is used to describe a variety of insurance products that pool investors and are structured such that the death of one member benefits the remaining living members. Tontines are ideal for individuals looking to insure against longevity risk, especially those lacking a beneficiary. Obviously, there is concern for such a product to maliciously take advantage of death, ultimately providing a few members with obscene and unfair profits. So tontines must be approached with care. For a brief history of tontines, see \cite{mckeever2009short}. For some ideas on the modern implementation and regulation of tontines, see \cite{milevsky2022managing}.

The pooled annuity funds considered here offer investors an increase in the probability to complete a schedule of withdrawals. In other words, the probability available through the pool is larger than what an individual can achieve alone. For the problem of maximizing this probability, only closed pooled annuity funds are considered. In a closed fund, members can only receive the scheduled withdrawals without any exceptions. 

Since this lack of liquidity is offputting to many investors, note that if living members are allowed to withdraw the present value of their initial contribution to the fund at any time, a remaining member's share of the combined funds is at least as large as that present value. This is because a remaining member's share takes into account the leftover contributions resulting from prior member deaths. A more detailed explanation of this logic, including how this present value is computed, is given in section \ref{s:conclusion}. Allowing a living member to withdraw the present value of their contribution can result in a probability of withdrawal success that differs from the closed pool case. However, this probability will still be larger than if the member had decided to invest alone instead of join the pool, all the while following the same portfolio weights as the pool.


\subsection{Literature Review}
Much of the research on pooled annuity funds uses a pool of annuitants that is homogeneous in contribution and mortality distribution. In the discrete-time setting with constant, deterministic returns and random times of death, consumption is set to be the amount a fair life annuity would pay if purchased with an annuitant's share of the current fund value \citep{bernhardt2021quantifying}. Then the relationship between pool size and stability of consumption over time is measured. In general, increasing the pool size offers a significant improvement to consumption stability. However, the improvement gained from each addition to the pool decreases as the pool size increases (i.e. consumption stability is a concave, increasing function of pool size). This positive relationship between pool size and consumption stability is supported by \cite{piggott2005simple}. In the continuous-time setting, where consumption and rebalancing is continuous in time, pooled annuity funds offer a significant increase to expected consumption \citep{stamos2008optimal}. There, consumption is optimized with respect to a (time-discounted) utility function supporting constant relative risk aversion. In contrast, the problem considered here fixes a schedule of consumption at time 0, and the goal is to optimize the probability of completing that schedule. 

The consumption stability of annuitants with different initial contributions to the fund is studied in \cite{bernhardt2023wealth}. The impact of allowing a new generation to enter the pool at each time step (for a finite number of time steps) is studied in \cite{donnelly2022practical}. Consumption stability is improved for earlier generations, but later generations face some instability. Instability results because their contributions are used to supplement earlier generations' consumption, and they cannot rely as heavily on the contributions of future generations to supplement their own consumption. The problem considered here is also concerned with consumption stability, but in a different way compared to the forementioned results. In particular, consumption amounts are fixed at time 0, and the goal is to maximize the probability of completing the consumption schedule. So stability here refers to whether or not the schedule is completed. 

The issue of actuarial fairness is addressed in \cite{donnelly2015actuarial}. In general, a pooled annuity fund is actuarial fair if all members recieve the same (relative) prospects in consumption. For example, in a pool of annuitants that is heterogeneous in age or contribution, the younger or poorer members should not have superior (relative) prospects in consumption. For heterogeneous pools, the number of members must be sufficiently large to minimize unfairness. In the homogeneous pool considered here, consumption increases with lifespan, but all members with a particular lifespan have the same consumption. So it is actuarially fair in this per-lifespan sense.

A mutual fund version of the pooled annuity fund has been proposed in \cite{goldsticker2007mutual}. While individuals have access to pooled annuity funds through particular employers \cite{comment2007mutual}, access is otherwise limited. Since pooled annuity funds can offer better deals compared to traditional annuities \cite{chen2023tontine}, it is worthwhile to consider making pooled annuity funds more accessible. For more information on this recent push to make pooled annuity funds more accessible, see \cite{milevsky2022managing}.

Use terminal wealth to refer to the combined funds remaining in a pooled annuity fund after all members have expired. A schedule of withdrawals can be completed if and only if the terminal wealth is non-negative. So maximizing the probability of completing a schedule of withdrawals is equivalent to maximizing the probability that terminal wealth is non-negative. Note that this is a version of the safety first principle, pioneered by \cite{roy1952safety}. Furthermore, this safety first maximization problem can be solved using dynamic programming. In particular, it is handled under a framework similar to the Borel setting of the discrete-time stochastic dynamic programming problem detailed in \citep{bertsekas1996stochastic}. Here, asset prices and portfolio weights are progressively measurable with respect to a filtration that represents the evolution of information over time. Moreover, the filtration can be continuous in time while the rebalancing and withdrawals are discrete in time.

\subsection{Summary of Main Results}
First the major assumptions are outlined. The pool is established at time 0, at which time each member contributes the same amount to the fund. The withdrawal schedule is determined at time 0, and it is the same for every member in the pool. Afterward, no additional individuals can join the pool, and no withdrawals can be made outside the predetermined withdrawal schedule. Assume the mortality distribution of each member in the pool is independent and identically distributed. 

Fix a sequence of rebalancing times so they are a superset of the withdrawal times. Using time-adapted portfolio weight functions, invest the combined funds in $n$ assets having independent increments in their log-returns (between rebalancing times). The probability for a particular member to complete the withdrawal schedule until death is maximized over those portfolio weight functions.

Applications use two assets, rebalanced on an annual basis: the S\&P Composite Index and an inflation protected bond. Only withdrawal schedules having equal annual withdrawals are considered. Note that the rebalancing and withdrawal times are set to coincide in this setting. The forementioned maximum probability is compared across various pool sizes, initial contributions and starting ages of members. In general, the maximum probability is an increasing function of pool size, initial contribution and starting age. The bulk of the (significant) benefit gained from joining a pool instead of investing alone is achieved with a surprisingly small pool, say 20 members. The additional benefit gained from joining a pool with more than 20 members is relatively small.

\subsection{Organization}
Section \ref{s:notation} provides the problem set-up. Theoretical results are given in section \ref{s:theory}. Proofs are omitted, since they are only slight modifications of the proofs given in \cite{brown}. Section \ref{s:app} provides applications of theoretical results using data. Data is described in section \ref{s:data}. Closing remarks and a discussion of related future research ideas are given in section \ref{s:conclusion}.

\section{Preliminaries}\label{s:notation}
Introduce the filtered probability space $(\Omega,\mathcal{F},\mathbb{F},\mathbb{P})$, where $\mathbb{F}:=\{\mathcal{F}(t)\}_{t\in T}$ denotes a filtration of $\mathcal{F}$ and $T\subset[0,\infty)$, $0\in T$. Consider $n$ assets available for investment, each denoted by an index from 1 to $n$. For each $j=1,2,...,n$, let $X_j:T\times\Omega\to(0,\infty)$ be an $\mathbb{F}$-adapted process. When convenient, write $X_j(t)$ in place of $X_j(t,\omega)$, understanding that $X_j(t)$ is an $\mathcal{F}(t)$-measurable function on $\Omega$. In this setting, $X_j(t)$ denotes the value of asset $j$ at time $t$. Let $\{t_k\}_{k=0}$ be an increasing sequence in $T$ with $t_0=0$. Require that for each $j$ and $t_k$, $\log X_j(t_{k+1})-\log X_j(t_k)$ is independent of $\mathcal{F}(t_k)$.  

Introduce a pool of $A_0$ annuitants with independent and identically distributed life distributions. By agreeing to be in this pool, each annuitant invests $P>0$ into a closed pooled annuity fund at time $0$. The fund is then invested in the forementioned $n$ assets, rebalancing only at those times $t_k$. If an annuitant is alive at time $t_k$, where $k=1,2,...$, then the annuitant withdraws $w_k\geq0$ from the fund at time $t_k$. No other withdrawals are allowed - not even on a dead annuitants behalf. Note that in this set-up, rebalancing can occur at time $t_k$ with $w_k=0$.

After accounting for $w_k$, denote the combined wealth in the fund at time $t_k$ with $W_k$. At each time $t_k$, rebalance $W_k$ according to the $\mathcal{F}(t_k)$-measurable portfolio weight vector $\boldsymbol\pi_k:\Omega\to\Pi$, where $\Pi=\{\mathbf{p}\in[0,1]^n:\sum_{j=1}^np_j=1\}$. When convenient, write $\boldsymbol\pi_k=(\pi_{k1},\pi_{k2},...,\pi_{kn})$, understanding that $\boldsymbol\pi_k$ and each $\pi_{kj}$ is an $\mathcal{F}(t_k)$-measurable function on $\Omega$. In particular, at each time $t_k$, invest $\pi_{kj}W_k$ in asset $j$ for each $j=1,2,...,n$. 

Track the number of annuitants as follows. Suppose an annuitant's status as alive or dead at time $t_{k+1}$ depends only on the annuitant's age and status at time $t_k$. Let $s$ denote the starting age of all annuitants, and let $d_k$ denote the probability of an annuitant aged $s+k$ at time $t_k$ dying by time $t_{k+1}$. For $i=1,2,...,A_0$ and $k=0,1,...$, define the $\mathcal{F}(t_{k+1})$-measurable random variable $B_k^i\sim\text{Bernoulli}(d_k)$. Require that each $B_k^i$ is independent of $\mathcal{F}(t_{k})$ and $X_j(t)$ for every $j=1,2,...,n$ and $t\in T$. Then the number of living annuitants at time $t_k$ is given by $A_k$, where 
\begin{equation}\label{eq:AB}
A_k=A_{k-1}-\sum_{i=1}^{A_{k-1}}B_{k-1}^i,\quad k=1,2,...
\end{equation}

To simplify notation, let $X_{jk}=X_j(t_{k+1})/X_j(t_{k})$ for $j=1,2,...,n$ and $k=0,1,...$. When convenient, write $\mathbf{X}_k=(X_{1k},X_{2k},...,X_{nk})$. Let $Y_{k}=\sum_{j=1}^n\pi_{kj}X_{jk}$ for $k=0,1,...$. Then the pooled wealth at time step $t_k$ is given by $W_k$, where the $W_k$ are computed recursively via
\begin{equation}
\begin{split}
W_0&=A_0P,\\
W_k&=Y_{k-1}W_{k-1}-A_kw_k,\quad k=1,2,...
\end{split}
\label{recursion}
\end{equation}
Again, note that wealth at time step $t_k$ is computed after accounting for the withdrawal of $w_k$ by all living annuitants at time step $t_k$. Furthermore, observe that each $X_{j,k-1}$, $Y_{k-1}$ and $W_k$ is an $\mathcal{F}(t_k)$-measurable function on $\Omega$. 

Observe that $W_k(\omega)<0$ implies $W_{k+1}(\omega)<0$ for each $\omega\in\Omega$. This guarantees that a failure to execute the scheduled withdrawals up to time $t_k$, indicated by $W_k(\omega)<0$, will be carried over to time $t_{k+1}$ and indicated by $W_{k+1}(\omega)<0$. 

Observe that $W_k$ is a function of each $\boldsymbol\pi_i$ for $i=0,1,...,k-1$. The notation
\begin{equation*}
\underset{\boldsymbol\pi_0,\boldsymbol\pi_1,...,\boldsymbol\pi_{k-1}}{\sup}\mathbb{P}(W_k\geq w)
\end{equation*}
is used to denote the supremum of $W_k$ over all $\mathcal{F}(t_i)$-measurable portfolio weight vectors $\boldsymbol\pi_i$, where $i=0,1,...,k-1$. This kind of abbreviation is used in similar situations where there is a $W_k$-like function that is constructed using the $\mathcal{F}(t_i)$-measurable $\boldsymbol\pi_i$. 

Use $\mathbb{E}[\ \cdot\ ]$ to denote the expectation with respect to $(\Omega,\mathcal{F},\mathbb{P})$. Denote the smallest $\sigma$-algebra containing the family of sets $S$ with $\sigma(S)$. Use $\mathbb{R}$ to denote the real numbers, and let $\mathbb{N}_0=\{0,1,2,...\}$. Given $u:\mathbb{R}\to\mathbb{R}$ and $Z:\Omega\to\mathbb{R}$, use $u(Z)$ to denote $u\circ Z$. Given sets $\Psi$, $I$ and $\{(f_i:\Psi\to\mathbb{R}):i\in I\}$, use $(\sup_{i\in I}f_i):\Psi\to\mathbb{R}$ to denote the pointwise supremum of the $f_i$, meaning for each $\psi\in\Psi$, $(\sup_{i\in I}f_i)(\psi)=\sup_{i\in I}(f_i(\psi))$. Let $\mathbf{1}=(1,1,...,1)$ denote the $n$-dimensional vector of 1s, and use $\cdot$ to indicate the dot product. 

\section{Theoretical Results}\label{s:theory}
Introduce the recursion starting with
\begin{equation*}
\widehat{W}_0=W_0,\quad I_0=1,
\end{equation*}
and for $k=1,2,...$,
\begin{equation}
\begin{split}
I_k&=I_{k-1}-\sum_{j=1}^{I_{k-1}}B_{k-1}^j,\\
\widehat{W}_k&=Y_{k-1}\widehat{W}_{k-1}-I_kA_kw_k.
\end{split}
\label{recursionhat}
\end{equation}
Here, $I_k$ is a flag indicating whether a particular annuitant is alive ($I_k=1$) or dead ($I_k=0$) at time $t_k$. In this setup, $\widehat{W}_k=W_k$ as long as $I_k=1$. If $I_{k}=0$, then $\widehat{W}_{k}\geq0$ iff $\widehat{W}_{k-1}\geq0$. In words, $\widehat{W}_k$ is non-negative if and only if the combined funds in the pool are not exhausted after executing the scheduled withdrawals for those times, up to and including $t_{k}$, at which the annuitant is alive.

Define $\tau:\Omega\to\mathbb{N}_0$ such that
\begin{equation*}
\tau(\omega)=\min\{k:I_k(\omega)=0\}.
\end{equation*} 
Then $\tau$ indicates the index of the first time step at which the annuitant has expired. The goal is to find
\begin{equation}\label{eq:argmax}
\underset{\boldsymbol\pi_0,\boldsymbol\pi_1,...,\boldsymbol\pi_{\tau-1}}{\sup}\mathbb{P}(\widehat{W}_{\tau}\geq 0). 
\end{equation}
In words, \eqref{eq:argmax} is the sumpremal probability of the annuitant completing the schedule of withdrawals until death, and the supremum is taken over the portfolio vectors $\boldsymbol\pi_i$, where $i=0,1,...,\tau-1$.

By the law of total probability,
\begin{equation}\label{eq:wtau0}
\mathbb{P}(\widehat{W}_{\tau}\geq 0)=\sum_{i=0}^{\infty}\mathbb{P}(\widehat{W}_i\geq 0,\ \tau=i).
\end{equation}
Observe from \eqref{recursionhat} that if $k\geq\tau(\omega)$, then 
\begin{equation*}
\widehat{W}_k\geq0\iff \widehat{W}_{\tau(\omega)}\geq0.
\end{equation*}
It follows that
\begin{equation}\label{eq:wtau0k}
\mathbb{P}(\widehat{W}_k\geq 0)=\mathbb{P}(\widehat{W}_k\geq 0,\ \tau>k)+\sum_{i=0}^{k}\mathbb{P}(\widehat{W}_i\geq 0,\ \tau=i).
\end{equation}
If $\mathbb{P}(\tau>k)=0$, then \eqref{eq:wtau0} and \eqref{eq:wtau0k} coincide. Alternatively, $k$ can be chosen large enough such that $\mathbb{P}(\tau>k)$ is sufficiently close to $0$, in which case \eqref{eq:wtau0} is approximated by \eqref{eq:wtau0k}. It follows that for $k$ sufficiently large,
\begin{equation}\label{eq:lswk}
\underset{\boldsymbol\pi_0,\boldsymbol\pi_1,...,\boldsymbol\pi_{k-1}}{\sup}\mathbb{P}(\widehat{W}_k\geq 0)\approx\underset{\boldsymbol\pi_0,\boldsymbol\pi_1,...,\boldsymbol\pi_{\tau-1}}{\sup}\mathbb{P}(\widehat{W}_{\tau}\geq 0).
\end{equation}
From here, the goal is to compute the left side of \eqref{eq:lswk}.

First define the function $v_k:\mathbb{R}\times\mathbb{N}_0\to[0,1]$ such that 
\begin{equation}\label{eq:vk}
v_k(x,a)=\begin{cases}1,&x\geq 0\\
0,&\text{otherwise}
\end{cases}
\end{equation}
Let $v_i:\mathbb{R}\times\mathbb{N}_0\to[0,1]$, $i=0,1,...,k-1$, denote the functions satisfying
\begin{equation}\label{eq:VI}
\begin{split}
v_i(x,a)&=(1-d_i)\max_{\mathbf{p}\in\Pi}\sum_{l=0}^{a-1}\binom{a-1}{l}(d_i)^l(1-d_i)^{a-1-l}h_{i+1}(x,a-l,\mathbf{p})\\&\quad+\ d_iv_k(x,0)\\
h_{i+1}(x,\widetilde{a},\mathbf{p})&=\mathbb{E}\big[v_{i+1}\big((\mathbf{p}\cdot\mathbf{X}_i)x-\widetilde{a}w_{i+1},\ \widetilde{a}\big)\big].
\end{split}
\end{equation}
For each $i=0,1,...,k-1$ and $a\in\mathbb{N}_0$, $v_i(x,a)$ is non-decreasing and upper semicontinuous over $x\in\mathbb{R}$. Moreover, the left side of \eqref{eq:lswk} is given by $v_0(W_0,A_0)$, which can be computed recursively, starting with \eqref{eq:vk} and then using \eqref{eq:VI}. Proofs of the previous statements are only slight modifications of the proofs given in \cite{brown} for the unpooled case, so they are left out. A simple inductive argument also shows that $v_i(x,a)=0$ for $i=0,1,...,k$, $x<0$ and $a\in\mathbb{N}_0$. Furthermore, $v_i(x,0)=v_k(x,0)$ for $i=0,1,...,k$ and $x\in\mathbb{R}$.

In a similar fashion, it is possible to maximize the probability for all annuitants to complete the schedule of withdrawals until death. First define $\widetilde{\tau}:\Omega\to\mathbb{N}_0$ such that
\begin{equation*}
\widetilde{\tau}(\omega)=\min\{k:A_k(\omega)=0\}.
\end{equation*} 
Then $\widetilde{\tau}$ indicates the index of the first time step at which all members of the pool have expired. The goal is to find
\begin{equation}\label{eq:argmax}
\underset{\boldsymbol\pi_0,\boldsymbol\pi_1,...,\boldsymbol\pi_{\widetilde{\tau}-1}}{\sup}\mathbb{P}(W_{\widetilde{\tau}}\geq 0). 
\end{equation}
In words, \eqref{eq:argmax} is the sumpremal probability for every annuitant to complete the schedule of withdrawals until death, and the supremum is taken over the portfolio vectors $\boldsymbol\pi_i$, where $i=0,1,...,\tau-1$. Like before, a sufficiently large $k$ gives
\begin{equation}\label{eq:argmaxtilde}
\underset{\boldsymbol\pi_0,\boldsymbol\pi_1,...,\boldsymbol\pi_{k-1}}{\sup}\mathbb{P}(W_k\geq 0)\approx\underset{\boldsymbol\pi_0,\boldsymbol\pi_1,...,\boldsymbol\pi_{\widetilde{\tau}-1}}{\sup}\mathbb{P}(W_{\widetilde{\tau}}\geq 0). 
\end{equation}

Using the same kind of logic as before, it follows that the left side of \eqref{eq:argmaxtilde} is given by $\widetilde{v}_0(W_0,A_0)$, where $\widetilde{v}_k:\mathbb{R}\times\mathbb{N}_0\to[0,1]$ is such that $\widetilde{v}_k=v_k$, and for $i=0,1,...,k-1$, $\widetilde{v}_i:\mathbb{R}\times\mathbb{N}_0\to[0,1]$ is such that
\begin{equation*}
\begin{split}
\widetilde{v}_i(x,a)&=\max_{\mathbf{p}\in\Pi}\sum_{l=0}^a\binom{a}{l}(d_i)^l(1-d_i)^{a-l}\widetilde{h}_{i+1}(x,a-l,\mathbf{p})\\
\widetilde{h}_{i+1}(x,\widetilde{a},\mathbf{p})&=\mathbb{E}\big[\widetilde{v}_{i+1}\big((\mathbf{p}\cdot\mathbf{X}_i)x-\widetilde{a}w_{i+1},\ \widetilde{a}\big)\big].
\end{split}
\end{equation*}

\subsection{Computing $v_0(W_0,A_0)$ with stock-bond portfolios}\label{s:sb}
Fix $T=[0,\infty)$ and $n=2$. Let $X_1(t)$ denote the value of the stock at time $t$. Assume the $X_{1i}$ are continuous in the sense that $\mathbb{P}(X_{1i}=x)=0$ for each $x>0$ and $i=0,1,...,k-1$. Let $X_2(t)=(1+r)^t$, meaning $X_2(t)$ denotes the value of the bond, with interest $r\geq0$, at time $t$. Let $\mathbb{F}$ be the natural filtration generated by $(X_1(t),X_2(t))$.

For $a\in\mathbb{N}_0$, let $m_{a,k}=0$, and for $i=0,1,...,k-1$, let 
\begin{equation}\label{eq:wir}
m_{a,i}=\frac{m_{a,i+1}+aw_{i+1}}{1+r}.
\end{equation}
Require that $w_k>0$, meaning there is a positive withdrawal at the last time step.

A simple inducetive argument shows that if $x\geq m_{a,i}$, then $v_i(x,a)=1$. Recall that in addition, $v_i(x,a)=0$ for $x<0$. Therefore
\begin{equation}\label{eq:hiR}
\begin{split}
h_{i+1}(x,a,\mathbf{p})=&\int_{R}v_{i+1}\big((\mathbf{p}\cdot\mathbf{X}_i)x-aw_{i+1},\ a\big)d\mathbb{P}\\
&+\ \mathbb{P}\big((\mathbf{p}\cdot\mathbf{X}_i)x-aw_{i+1}\geq m_{a,i+1}\big),
\end{split}
\end{equation}
where $R=\{\omega:0\leq(\mathbf{p}\cdot\mathbf{X}_i)x-\widetilde{a}w_{i+1}<m_{a,i+1}\}$.

The recursion can also be kick-started at $k-1$, since for $q\neq0$,
\begin{equation}\label{eq:vkm1}
\begin{split}
h_k(x,a,(q,1-q))&=\max_{\mathbf{p}\in\Pi}\mathbb{P}(((q,1-q)\cdot\mathbf{X}_i)x-aw_k\geq 0)\\
&=\mathbb{P}\Big(\frac{X_{1,k-1}}{1+r}\geq 1+\frac{1}{q}\big(\frac{m_{a,k-1}}{x}-1\big)\Big).
\end{split}
\end{equation}

\section{Applications}\label{s:app}
Like in \cite{brown}, theoretical results are applied in the case where $n=2$. The two assets used are the S\&P Composite Index and an inflation-protected bond. The annual real returns of the S\&P Composite Index appear to be historically stable in the sense that they are approximately iid (Normal with mean 1.083 and standard deviation .1753) over the past 150 years \citep{brown}. For $n>2$ the optimization is more difficult to compute, and it is not easy to find other assets like the S\&P Composite Index that have such historical stability.

To construct the mortality distribution of annuitants, applications use the 2017 per-age death rates of the US Social Security area population. Section \ref{s:data} describes the S\&P Composite Index data and the mortality distribution data. Section \ref{s:appSetup} details the set-up needed to apply theoretical results and describes the algorithms used in applications. Results of applications are given in section \ref{s:resultsA}.

\subsection{Data}\label{s:data}
Annual data from the S\&P Composite Index and Comsumer Price Index is taken from \url{http://www.econ.yale.edu/~shiller/data.htm}, collected for easy access at \url{https://github.com/HaydenBrown/Investing}. The data spans 1871 to 2020 and is described in table \ref{t:data}. Note that S\&P Composite Index refers to Cowles and Associates from 1871 to 1926, Standard \& Poor 90 from 1926 to 1957 and Standard \& Poor 500 from 1957 to 2020. Cowles and Associates and the S\&P 90 are used here as backward extensions of the S\&P 500.

\begin{figure}
  \includegraphics[width=\linewidth]{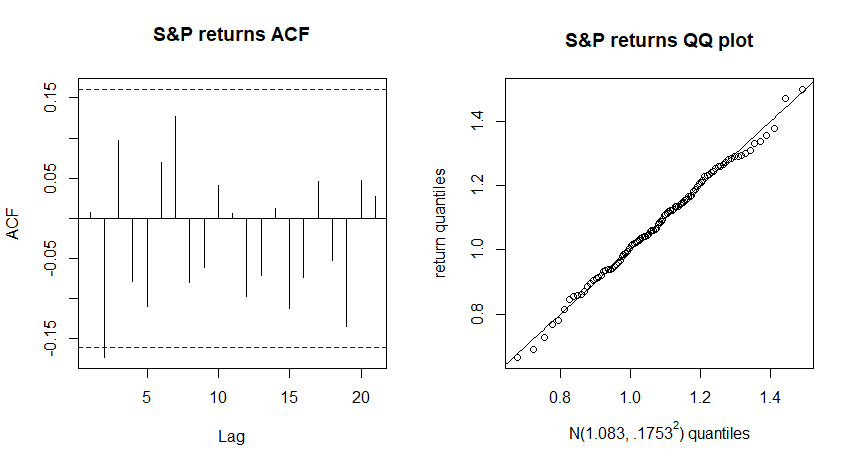}
  \caption{Left: The autocorrelation function of annual S\&P log-returns. Right: Quantiles of annual S\&P returns versus $\mathcal{N}(1.083,.1753^2)$ quantiles.}
  \label{fig:acfqq}
\end{figure}

The data is transformed so that annual returns incorporate dividends and are adjusted for inflation. In particular, returns are computed using the consumer price index, the S\&P Composite Index price and the S\&P Composite Index dividend. Use the subscript $k$ to denote the $k$th year of $C$, $I$ and $D$ from Table \ref{t:data}. The return for year $k$ is computed as $\frac{I_{k+1}+D_k}{I_k}\cdot\frac{C_k}{C_{k+1}}$. The justification for treating S\&P returns as independent and identically distributed Normal random variables with mean 1.083 and standard deviation .1753 is given in \cite{brown}. To give a brief visual of this justification, Figure \ref{fig:acfqq} shows the autocorrelation function of sample returns and how the sample quantiles align with the Normal quantiles.

\begin{table}
\begin{center}
\caption{Data variable descriptions.}\label{t:data}
\begin{tabular}{ ll } 
\toprule
Notation & Description \\
\toprule
 I & Average monthly close of the S\&P composite index \\ 
 D & Dividend per share of the S\&P composite index \\ 
 C & January consumer price index \\ 
\toprule
\end{tabular}
\end{center}
\end{table}

\begin{figure}
  \includegraphics[width=\linewidth]{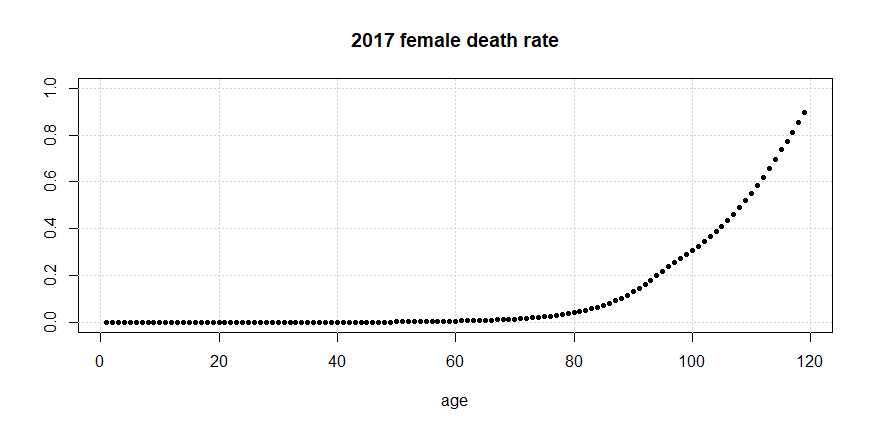}
  \caption{2017 female death rates for the US Social Security area population.}
  \label{fig:fdr}
\end{figure}

Death rates are taken from \url{https://www.ssa.gov}, the official website of the Social Security Administration. In particular, the female per-age death rates of the US Social Security area population are taken from the 2017 period life table. Female death rates are used because they are generally lower than male death rates. The female death rates are illustrated in figure \ref{fig:fdr}.
Let $d_j$ denote the 2017 female death rate for age $j$, and let $s$ denote the starting age for a given schedule of investments and withdrawals. Applications use $k=120-s$ and
\begin{equation*}
p_i=d_{s+i},\quad i=0,1,...,k-1.
\end{equation*}

\subsection{Set-up}\label{s:appSetup}
In order to apply theoretical results, $T$, $n$, $X_j(t)$ for $j=1,2,...,n$, $\mathbb{F}$ and $\{t_k\}_{k=0}$ need to be specified. Set $T=[0,\infty)$ and $n=2$. Let $X_1(t)$ denote the inflation-adjusted value of the S\&P Composite Index at time $t$, and let $X_2(t)=(1+r)^t$, meaning $X_2(t)$ denotes the inflation-adjusted value of an inflation-protected bond, with interest $r$, at time $t$. Set $t_k=k$ for $k=0,1,...$. Let $\{\mathcal{G}(t)\}_{t\in T}$ be the natural filtration generated by $(X_1(t),X_2(t))$. Then define $\mathbb{F}$ as follows.
\begin{equation*}
\mathcal{F}(t):=\sigma\Big(\mathcal{G}(t)\cup\underset{i=1,2,...,A_0}{\bigcup_{k-1\leq t}}\sigma(B_k^i)\Big),\quad t\in T.
\end{equation*}

Since $X_1(t)$ and $X_2(t)$ are inflation-adjusted, it follows that the $w_k$ and $W_k$ are also inflation-adjusted. For example, if $w_k=2$ and inflation is 5\% from time 0 to time $t_k$, then the actual amount withdrawn at time $t_k$ is $2\cdot 1.05$. In other words, 2 is the inflation-adjusted amount withdrawn, and $2\cdot 1.05$ is the actual amount withdrawn. 

Theoretical results also require the $\mathbf{X}_k$ to be independent of $\mathcal{F}(t_k)$. It suffices to have iid $\mathbf{X}_k$. The treatment of S\&P returns as iid has already been addressed, and the returns of the inflation protected bond are clearly iid because they are deterministic. 

\subsubsection{Computing $v_0(W_0,A_0)$}\label{s:valg} 
Let $M$ be a sufficiently large postive integer. In algorithm \ref{alg:v}, $v_i(x,a)$ is computed recursively for $x\in D_{a,i}$, where
\begin{equation*}
D_{a,i}=\Big\{\frac{jm_{a,i}}{M}:j=1,...,M-1\Big\}.
\end{equation*}
The following elaborates on the details behind algorithm \ref{alg:v}. 

Recall that $v_k$ is given by \eqref{eq:vk}. Observe that the set-up detailed at the top of section \ref{s:appSetup} aligns with that of section \ref{s:sb}. So $v_i(x,a)=1$ for $x\geq m_{a,i}$ and $i=0,1,...,k$. Furthermore, $v_{k-1}(x,a)$ can be computed using \eqref{eq:vkm1} for $x\in(0,m_{a,k-1})$.

Denote the pdf and cdf of the iid $X_{1i}$ with $f$ and $F$, respectively. Then \eqref{eq:hiR} implies that for each $x\in(0,m_{a,i})$ and $q\neq0$, $h_i(x,a,(q,1-q))$ is given by
\begin{equation}\label{eq:vqf}
\int_b^{b+\frac{m_{a,i}}{qx}}v_i((qz+(1-q)(1+r))x-aw_i,a)f(z)dz+1-F\Big(b+\frac{m_{a,i}}{qx}\Big),
\end{equation}
where $q$ indicates the proportion invested in the stock at time $t_{i-1}$ and
\begin{equation*}
b=1+r-\frac{1+r}{q}+\frac{aw_i}{qx}.
\end{equation*}
Transforming the integral in \eqref{eq:vqf} with the substitution 
\begin{equation*}
y=(qz+(1-q)(1+r))x-aw_i
\end{equation*} 
yields
\begin{equation}\label{eq:vqfT}
\frac{1}{qx}\int_0^{m_{a,i}} v_i(y,a)f\Big(1+r-\frac{1+r}{q}+\frac{y+aw_i}{qx}\Big)dy.
\end{equation}
Algorithm \ref{alg:v} approximates \eqref{eq:vqfT} with
\begin{equation}\label{eq:vqfTS}
\Bigg(F\Big(b+\frac{m_{a,i}}{qx}\Big)-F(b)\Bigg)\cdot\frac{\sum_{y\in D_{a,i}}v_{i}(y,a)f\Big(1+r-\frac{1+r}{q}+\frac{y+aw_i}{qx}\Big)}{\sum_{y\in D_{a,i}}f\Big(1+r-\frac{1+r}{q}+\frac{y+aw_i}{qx}\Big)}.
\end{equation}
Note that the fraction (right side) in \eqref{eq:vqfTS} approximates the expectation of $v_i(Y,a)$ given $0<Y<m_{a,i}$, where $Y$ has pdf
\begin{equation*}
\frac{1}{qx}\cdot f\Big(1+r-\frac{1+r}{q}+\frac{y+aw_i}{qx}\Big).
\end{equation*}
Furthermore, $\mathbb{P}(0<Y<m_{a,i})$ is given by the big parenthesis in \eqref{eq:vqfTS}. Summarizing, for $x\in(0,m_{a,i})$ and $q\neq0$, algorithm \ref{alg:v} approximates $h_i(x,a,(q,1-q))$ with 
\begin{equation*}
\eqref{eq:vqfTS}+1-F\Big(b+\frac{m_{a,i}}{qx}\Big).
\end{equation*}

Now recall from \eqref{eq:VI} that $v_i(x,a)=\max_{q\in[0,1]}g_{i+1}(x,a,q)$, where
\begin{equation}\label{eq:maxqh}
\begin{split}
g_{i+1}(x,a,q)&=(1-d_i)\sum_{l=0}^{a-1}\binom{a-1}{l}(d_i)^l(1-d_i)^{a-1-l}h_{i+1}(x,a-l,(q,1-q))\\
&\quad+\ d_iv_k(x,0).
\end{split}
\end{equation}
In algorithm \ref{alg:v}, $\max_{q\in[0,1]}g_{i+1}(x,a,q)$ is computed using an iterated grid search, where the grid is refined at each iteration. In particular, the first grid tests $q$ in $G_1=\{.1,.2,...,.9\}$. Let $q_1$ denote the $q$ in $G_1$ that produces the maximum of $g_{i+1}(x,a,q)$. The next grid is $G_2=\{q_1\pm .01j: j=-9,-8,...,10\}$. Let $q_2$ denote the $q$ in $G_2$ that produces the maximum of $g_{i+1}(x,a,q)$. From here, algorithm \ref{alg:v} uses the approximation
\begin{equation*}
\max_{q\in(0,1]}g_{i+1}(x,a,q)\approx g_{i+1}(x,a,q_2).
\end{equation*}
Then algorithm \ref{alg:v} compares said approximation with $g_{i+1}(x,a,0)$ to approximate $\max_{q\in[0,1]}g_{i+1}(x,a,q)$. 

$g_{i+1}(x,a,0)$ is computed as follows. Observe that 
\begin{equation}\label{eq:gi0}
\begin{split}
g_{i+1}(x,a,0)&=(1-d_i)\sum_{l=0}^{a-1}\Bigg[\binom{a-1}{l}(d_i)^l(1-d_i)^{a-1-l}\\
&\quad\quad\quad\quad\quad\quad\quad\cdot v_{i+1}\big((1+r)x-(a-l)w_{i+1},\ a-l\big)\Bigg]\\
&\quad+\ d_iv_k(x,0).
\end{split}
\end{equation}
Let $\theta=(1+r)x-(a-l)w_{i+1}$. Algorithm \ref{alg:v} approximates each $v_{i+1}(\theta,a-l)$ in \eqref{eq:gi0} with the following lower bound:
\begin{equation*}
v_{i+1}(\theta,a-l)\approx\begin{cases}
0&\theta<\frac{m_{a-l,i+1}}{M}\\
1&\theta\geq m_{a-l,i+1}\\
\underset{y\in D_{a-l,i+1}\cap[0,\theta]}{\max}v_{i+1}(y,a-l)&\text{otherwise}.
\end{cases}
\end{equation*}

\subsubsection{Simulating $\mathbb{P}(\widehat{W}_k\geq 0)$} \label{s:usim}
Algorithm \ref{alg:usim} computes $\mathbb{P}(\widehat{W}_k\geq 0)$ via simulation. It uses $\boldsymbol\pi_{i}$ that are defined in the following way using Borel measurable $q_i:\mathbb{R}\times\{0,1,...,A_0\}\to[0,1]$.
\begin{equation*}
\boldsymbol\pi_{i}:=(q_i(\widehat{W}_i,A_i),1-q_i(\widehat{W}_i,A_i)),\quad i=0,1,...,k-1.
\end{equation*}
First, $N$ realizations of $\widehat{W}_k$ are simulated. Then $\mathbb{P}(\widehat{W}_k\geq 0)$ is computed as the number of non-negative realizations, divided by $N$. 

When simulating $\mathbb{P}(\widehat{W}_k\geq 0)$ with the optimal portfolio weight functions, do the following. Execute algorithm \ref{alg:v} and return the $q^*_i(x,a)$ for $x\in D_{a,i}$ and $a\in\{0,1,...,A_0\}$. The values of $q^*_i(x,a)$ for $x\notin D_{a,i}$ are computed via linear interpolation. Set 
\begin{equation*}
q^*_i(x,a)=\begin{cases}
1&x\leq0\\
0&x\geq m_{a,i}.
\end{cases}
\end{equation*}
For $x\in (0,m_{a,i})\setminus D_{a,i}$, let $y_x$ denote the largest element of $D_{a,i}\cup\{0\}$ that is less than or equal to $x$, and set
\begin{equation*}
q^*_i(x,a)=q^*_i(y_x,a)+\frac{x-y_x}{m_{a,i}/M}\cdot(q^*_i(y_x+m_{a,i}/M,a)-q^*_i(y_x,a)).
\end{equation*}
To simulate the maximum $\mathbb{P}(\widehat{W}_k\geq 0)$ simply follow algorithm \ref{alg:usim} using this filled-in $q^*_i$ in place of $q_i$.

\subsection{Results}\label{s:resultsA}
First, $v_0(ax,a)$ is computed using algorithm \ref{alg:v} with $r=0$, $M=100$, $s=65$ and $X_{1i}\sim\mathcal{N}(1.083,.1753^2)$ and $w_{i+1}=1$ for $i=0,1,2,...,k-1$. The interest rate $r$ is set at 0 because returns are already inflation adjusted, and any additional interest obtained from an inflation protected bond is likely to be low. After experimenting with different values of $M$, 100 appears to produce accurate results in a reasonable amount of time. For example, computing $v_0(ax,a)$ for $a=1,2,...,100$ takes under ten hours. A starting age of 65 is selected because it is a common age to begin retirement. Equal annual withdrawals of one unit are chosen to focus on annuitants looking to have a constant and reliable income until death.

\begin{figure}
  \includegraphics[width=\linewidth]{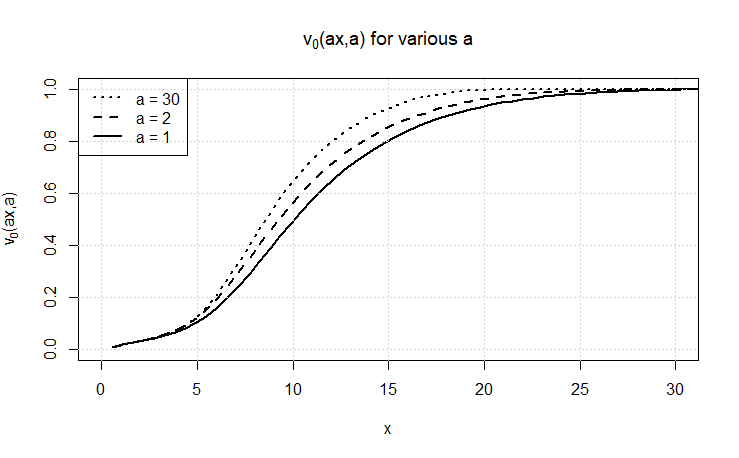}
  \caption{Using algorithm \ref{alg:v} with $r=0$ and $M=100$, illustrates $v_0(ax,a)$ over $x$ for a starting age of $65$ (i.e. $s=65$) and various $a$. Note the assumption that $X_{1i}\sim\mathcal{N}(1.083,.1753^2)$ and $w_{i+1}=1$ for $i=0,1,2,...,k-1$.}
  \label{fig:vdifa}
\end{figure}

\begin{figure}
  \includegraphics[width=\linewidth]{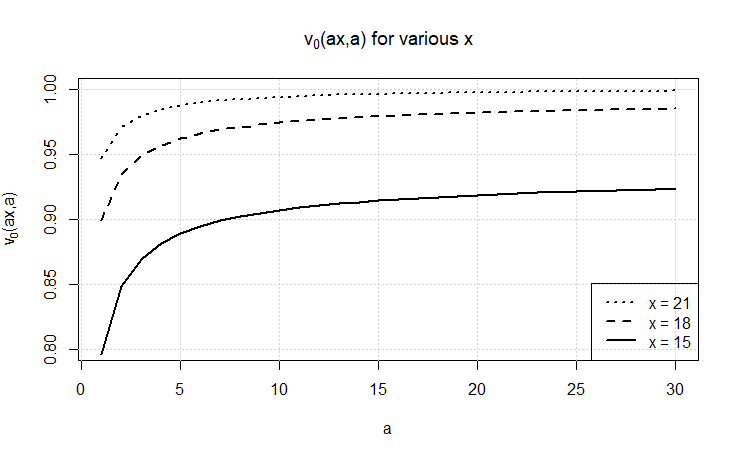}
  \caption{Using algorithm \ref{alg:v} with $r=0$ and $M=100$, illustrates $v_0(ax,a)$ over $a$ for a starting age of $65$ (i.e. $s=65$) and various $x$. Note the assumption that $X_{1i}\sim\mathcal{N}(1.083,.1753^2)$ and $w_{i+1}=1$ for $i=0,1,2,...,k-1$.}
  \label{fig:vdifx}
\end{figure}

Figures \ref{fig:vdifa} and \ref{fig:vdifx} show how $v_0(ax,a)$ changes over $a$ and $x$. In general, $v_0(ax,a)$ increases as $a$ or $x$ increases. $v_0(ax,a)$ is also concave with respect to $a$. Looking at figure \ref{fig:vdifa}, observe that a noticeable increase in $v_0(ax,a)$ is obtained with just two annuitants instead of one. The two annuitant case could be implemented without an insurance company, between two friends or family members who are close in age. Looking at figure \ref{fig:vdifx}, 3 (7) annuitants can achieve 95\% (90\%) confidence in withdrawing $5.\bar{5}\%$ ($6.\bar{6}\%$) of the initial contribution, annually, until death.

\begin{figure}
  \includegraphics[width=\linewidth]{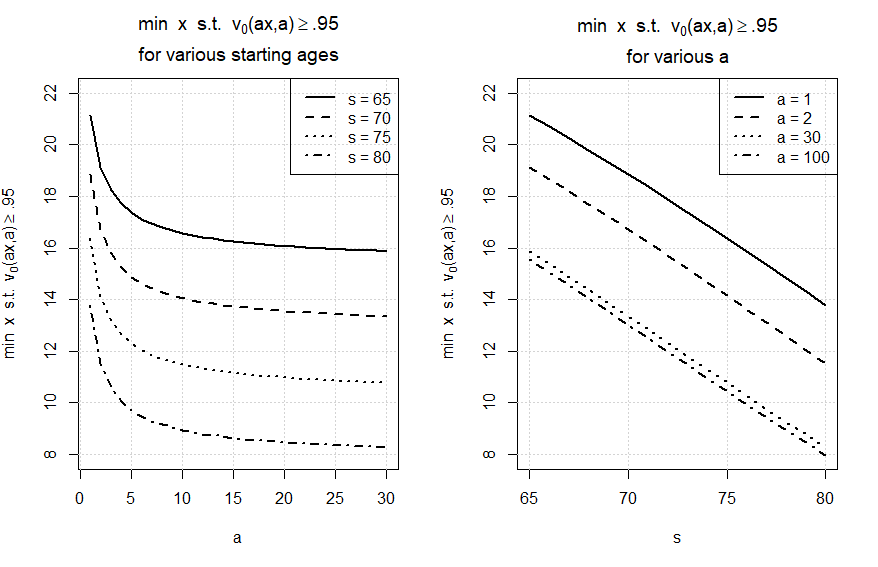}
  \caption{Using algorithm \ref{alg:v} with $r=0$ and $M=100$, illustrates the necessary and sufficient per-annuitant initial investment ($P=x$) to complete the withdrawal schedule with confidence $.95$ for various starting ages ($s$) and initial pool sizes ($A_0=a$). Note the assumption that $X_{1i}\sim\mathcal{N}(1.083,.1753^2)$ and $w_{i+1}=1$ for $i=0,1,2,...,k-1$.}
  \label{fig:vCdifasa}
\end{figure}

Figure \ref{fig:vCdifasa} illustrates the necessary and sufficient per-annuitant initial investment ($P=x$) to complete the forementioned withdrawal schedule with confidence $.95$ for various starting ages ($s$) and initial pool sizes ($A_0=a$). Call this necessary and sufficient initial investment $P^*$. When holding $A_0$ constant, $P^*$ decreases by roughly 2 to 2.5 for each 5 year increase in starting age. Increasing $A_0$ from 1 to 2 decreases $P^*$ by about 2 regardless of starting age. In general the decrease in $P^*$ resulting from an increase in $A_0$ is constant over various starting ages. Not much of a decrease in $P^*$ is available from increasing $A_0$ over 30. 

\begin{figure}
  \includegraphics[width=\linewidth]{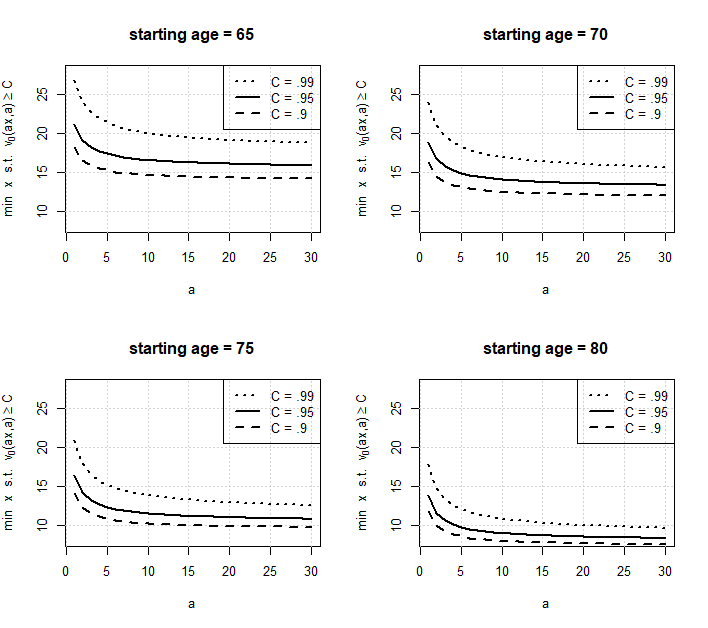}
  \caption{Using algorithm \ref{alg:v} with $r=0$ and $M=100$, illustrates the necessary and sufficient per-annuitant initial investment ($P=x$) to complete the withdrawal schedule with confidence $C$, for various starting ages ($s$) and initial pool sizes ($A_0=a$). Note the assumption that $X_{1i}\sim\mathcal{N}(1.083,.1753^2)$ and $w_{i+1}=1$ for $i=0,1,2,...,k-1$.}
  \label{fig:vCdifsa}
\end{figure}

Figure \ref{fig:vCdifsa} shows how $P^*$ changes when the confidence level is varied between .9, .95 and .99. In general, the magnitude of this change is larger for smaller $A_0$. In terms of $P^*$, 99\% confidence costs quite a bit more than 95\% confidence, and the extra 4\$ of confidence may not be worth the additional cost.

\begin{figure}
  \includegraphics[width=\linewidth]{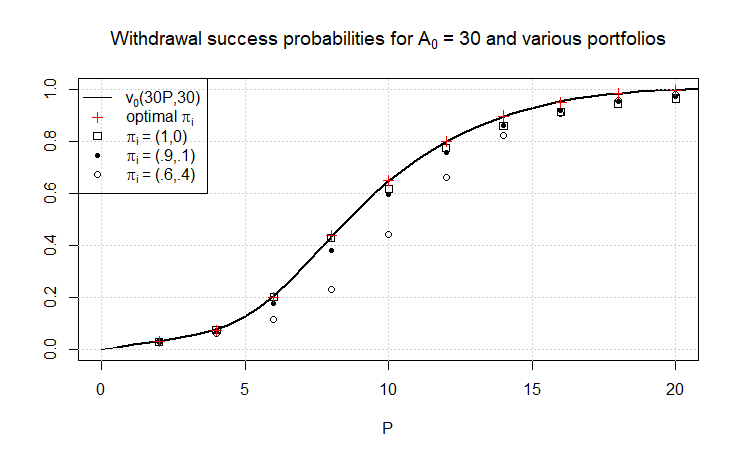}
  \caption{Using algorithm \ref{alg:v} with $r=0$ and $M=100$, illustrates the withdrawal success probabilities for an initial pool size of 30 annuitants, all aged $65$ (i.e. $s=65$). Recall that $P$ is the initial contribution of each annuitant. $v_0(30P,30)$ is produced using algorithm \ref{alg:v}. The individual points are all produced via simulation algorithm... Note the assumption that $X_{1i}\sim\mathcal{N}(1.083,.1753^2)$ and $w_{i+1}=1$ for $i=0,1,2,...,k-1$.}
  \label{fig:30plus}
\end{figure}

Figure \ref{fig:30plus} shows how $v_0(A_0P,A_0)$ compares with $\mathbb{P}(\widehat{W}_k\geq0)$ for various constant portfolio weight function, using $A_0=30$ and a starting age of 65. In general going all-in on the S\&P Composite Index gives the closest $\mathbb{P}(\widehat{W}_k\geq0)$ to $v_0(A_0P,A_0)$. However, the difference is noticeable (as much as .04) for $10\leq P\leq 20$. 

\begin{figure}
  \includegraphics[width=\linewidth]{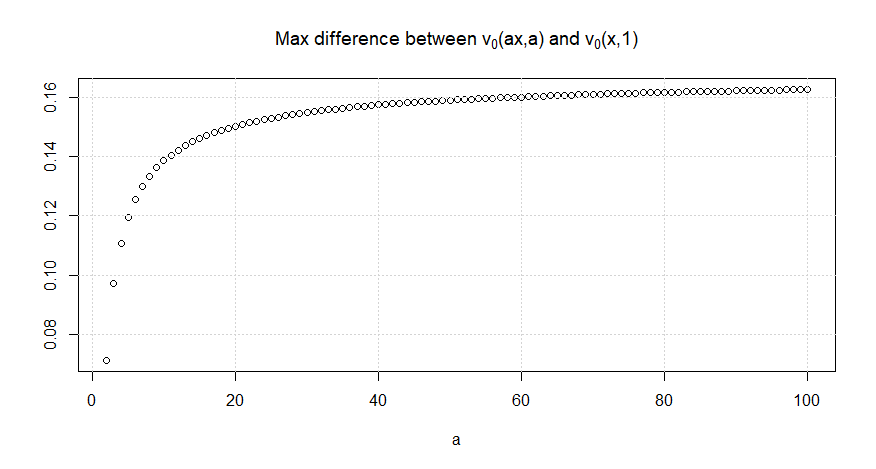}
  \caption{Using algorithm \ref{alg:v} with $r=0$ and $M=100$, illustrates 
\begin{equation}\label{eq:maxdifa1}
\max_{x>0}\{v_0(ax,a)-v_0(x,1)\}.
\end{equation}
for a starting age of $65$ (i.e. $s=65$) and $a=2,3,...,100$. Note the assumption that $X_{1i}\sim\mathcal{N}(1.083,.1753^2)$ and $w_{i+1}=1$ for $i=0,1,2,...,k-1$.}
  \label{fig:maxdifa1}
\end{figure}

Figure \ref{fig:maxdifa1} illustrates the maximum difference (with respect to $x$) between $v_0(ax,a)$ and $v_0(x,1)$ for $a$ from 1 to 100. The effect of pooling on withdrawal success probability is clearly significant. In particular, a small pool (10 to 20 annuitants) can achieve a withdrawal success probability that is .14 to .15 higher than what an individual can achieve alone. Larger pools improve on this maximum difference, but not by much. 

\begin{figure}
  \includegraphics[width=\linewidth]{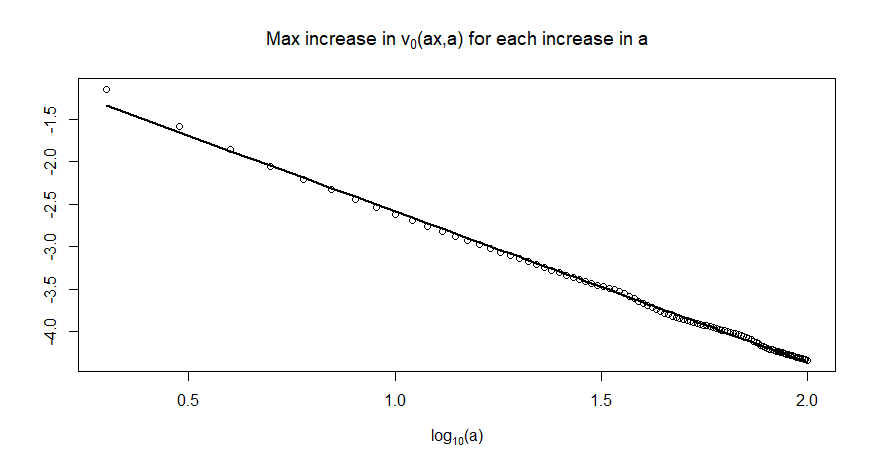}
  \caption{Using algorithm \ref{alg:v} with $r=0$ and $M=100$, illustrates 
\begin{equation}\label{eq:maxdif}
\log_{10}\max_{x>0}\{v_0(ax,a)-v_0((a-1)x,a-1)\}.
\end{equation}
for a starting age of $65$ (i.e. $s=65$) and $a=2,3,...,100$. Note the assumption that $X_{1i}\sim\mathcal{N}(1.083,.1753^2)$ and $w_{i+1}=1$ for $i=0,1,2,...,k-1$. The fitted line follows 
\begin{equation*}
-.805-1.78\log_{10}a.
\end{equation*}}
  \label{fig:vchange2}
\end{figure}

Following the apparrent trend in figure \ref{fig:vchange2}, for $a>a_0$ there is
\begin{equation*}
v_0(ax,a)\lessapprox v_0(a_0x,a_0)+\sum_{\widetilde{a}=a_0+1}^a10^{-.805}\cdot \widetilde{a}^{-1.78}.
\end{equation*}
Furthermore, observe that 
\begin{equation*}
\sum_{\widetilde{a}=a_0+1}^a\widetilde{a}^{-1.78}\leq\int_{a_0}^a\widetilde{a}^{-1.78}d\widetilde{a}\leq\frac{1}{.78\cdot a_0^{.78}}.
\end{equation*}
So it appears that when $a_0$ is sufficiently large, $v_0(ax,a)$ can only be slightly larger than $v_0(a_0x,a_0)$ when $a>a_0$. To give an idea of how large $a_0$ should be to avoid missing out on a potentially large increase in $v_0(ax,a)$ over $v_0(a_0x,a_0)$, see table \ref{t:ubsa}. 
\begin{table}
\begin{center}
\caption{Upper bounds for $\sum_{\widetilde{a}=a_0+1}^a10^{-.805}\cdot \widetilde{a}^{-1.78}$. These are also approximate upper bounds for $v_0(ax,a)-v_0(a_0x,a_0)$, where $a>a_0$ and $x>0$. }\label{t:ubsa}
\begin{tabular}{ rc } 
\toprule
$a_0$ & $10^{-.805}\cdot a_0^{-.78}/.78$\\
\toprule
100&.0055\\
250&.0027\\ 
500&.0016\\
1000&.0009\\
2000&.0005\\
4000&.0003\\
\toprule
\end{tabular}
\end{center}
\end{table}

\begin{figure}
  \includegraphics[width=\linewidth]{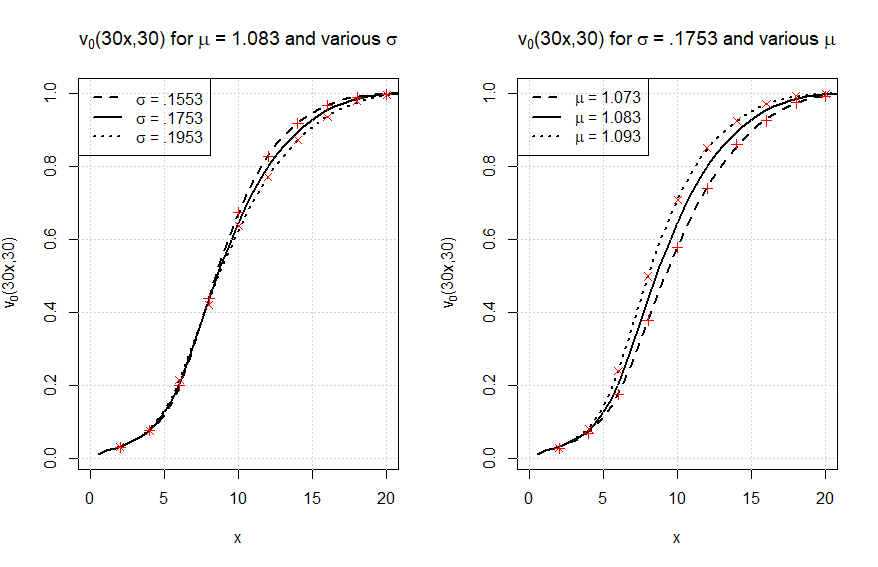}
  \caption{Using algorithm \ref{alg:v} with $r=0$, $M=100$ and $s=65$, illustrates $v_0(30x,30)$ for various combinations of $\mu$ and $\sigma$. Note the assumption that $X_{1i}\sim\mathcal{N}(\mu,\sigma^2)$ and $w_{i+1}=1$ for $i=0,1,2,...,k-1$. Let $\boldsymbol\pi_i^*$ denote the $\boldsymbol\pi_i$ returned from algorithm \ref{alg:v} using $X_{1i}\sim\mathcal{N}(1.083,.1753^2)$. Left: The red + and $\times$ indicate simulated versions of $\mathbb{P}(\widehat{W}_k\geq0)$ (from algorithm \ref{alg:usim}) when $\boldsymbol\pi_i=\boldsymbol\pi_i^*$ and $X_{1i}\sim\mathcal{N}(\mu,\sigma^2)$ with $(\mu,\sigma)=(1.083,.1553)$ and $(\mu,\sigma)=(1.083,.1953)$, respectively. Right: The red $\times$ and + indicate simulated versions of $\mathbb{P}(\widehat{W}_k\geq0)$ (from algorithm \ref{alg:usim}) when $\boldsymbol\pi_i=\boldsymbol\pi_i^*$ and $X_{1i}\sim\mathcal{N}(\mu,\sigma^2)$ with $(\mu,\sigma)=(1.093,.1753)$ and $(\mu,\sigma)=(1.073,.1753)$, respectively.}
  \label{fig:musig}
\end{figure}

Figure \ref{fig:musig} shows how changes to $\mu$ and $\sigma$ affect the optimal withdrawal success probability, provided $X_{1i}\sim\mathcal{N}(\mu,\sigma^2)$. Figure \ref{fig:musig} also shows that the maximum withdrawal success probability for $X_{1i}\sim\mathcal{N}(\mu,\sigma^2)$ can be nearly achieved using the optimal portfolio weights coming from Normal $X_{1i}$ with a slightly different mean and variance. So knowing the approximate distribution of the $X_{1i}$ appears to be sufficient to achieve a withdrawal success probability that is nearly optimal. 

\section{Conclusion}\label{s:conclusion}
Applications show that there is a worthwhile benefit to be gained from joining a pooled annuity fund instead of trying to complete a schedule of withdrawals independently. Moreover, the size of the pool does not have to be very large to reap most of the benefit available. If such a pooled annuity fund ever becomes available to the public, the provider will only need a small number of individuals, say 20, to establish an attractive pool. 

Here, only a single contribution, followed by a schedule of withdrawals, is studied. Note that it is also possible to study multiple contributions, occurring at different times, preceding the schedule of withdrawals. For example, each annuitant could make the same annual contribution for 20 years, and then the withdrawal schedule could start after that. This falls more in line with how pensions are structured. However, the heterogeneity of employees contributing to the same pension fund can complicate the kind of analysis conducted here. The emphasis placed here on a single contribution at time 0 is because it appeals directly to those retirement-age individuals looking to insure against longevity risk. The multiple contribution case appeals more to younger individuals planning for retirement.

In applications, the benefit of using optimal instead of constant portfolio weights was demonstrated for a special case. Remaining fully invested in the S\&P Composite Index gave a withdrawal success probability that was close to the optimum. On one hand, the difference between the two can be noticeable - as much as .04. On the other hand, this difference is relatively small and may not be considered worth the extra effort that comes with implementing the optimal portfolio weights. Out of curiosity, the author tested the effect of changing the mean and standard deviation of S\&P annual returns on this apparent closeness in the success probabilities between the portfolio that remains fully invested in the S\&P Composite Index and the optimal portfolio. The closeness remained. So if it is difficult to get annuitants to buy in to this idea of optimal portfolio weights, fund providers can rest somewhat easily knowing that annuitants will not miss out on much by investing in just the S\&P Composite Index.

One downside of the pooled annuity funds considered here is that they are closed. This lack of liquidity could easily turn away an interested individual. If living members are allowed the option of withdrawing all of the present value of their initial contribution from the combined funds, at any time, then this can change the probability of withdrawal success for the members who choose to remain in the pool until death. Future research could study the effect of this option on the success probability for those members who choose to remain in the pool until death. Regardless of whether this option is allowed, the success probability for an individual who remains in a pool until death is at least as high as if the individual invests independently, but follows the same portfolio weights as the pool. So if the pool is going all-in on the S\&P Composite Index, and this option is allowed, the success probability for an individual who remains in a pool until death is at least as high as if the individual invests independently in just the S\&P Composite Index. The logic is as follows.

At time $t_k$, the present value of a living member's contribution to the combined funds is given by $\max\{0,\widetilde{W}_k\}$, where 
\begin{equation*}
\begin{split}
\widetilde{W}_0&=P,\\
\widetilde{W}_k&=Y_{k-1}\widetilde{W}_{k-1}-w_k,\quad k=1,2,...
\end{split}
\end{equation*}
For $t\in(t_k,t_{k+1})\cap T$, the present value is 
\begin{equation*}
\max\{0,\widetilde{W}_k\}\cdot\sum_{j=1}^n\pi_{kj}X_{j}(t)/X_j(t_k).
\end{equation*}
Use $\overline{A}_k$ to denote the number of living members in the pool at time $t_k$, where members can only join the pool at time 0, but they can leave the pool by either dying or exercising the forementioned option. Use $\overline{W}_k$ to denote the resulting value of the combined funds at time $t_k$. Then since at most $A_{k-1}-A_{k}$ members can exercise the option between times $t_{k-1}$ and $t_k$, 
\begin{equation*}
\overline{W}_k\geq Y_{k-1}\big(\overline{W}_{k-1}-(A_{k-1}-A_{k})\max\{0,\widetilde{W}_{k-1}\}\big)-\overline{A}_kw_k.
\end{equation*}
An inductive argument reveals that $\overline{W}_k\geq A_k\widetilde{W}_k$ whenever $\widetilde{W}_k\geq0$. So 
\begin{equation*}
\widetilde{W}_k\geq0\implies\overline{W}_k\geq0.
\end{equation*}
This ultimately means that if an individual can complete the withdrawal schedule alone, then that individual could have completed the withdrawal schedule as a member of a pool instead, even if the pool allows living members to withdraw all of the present value of their initial contribution to the combined funds at any time. 

\begin{algorithm}
\caption{Compute $\mathbb{P}(\widehat{W}_k\geq w)$ given $\boldsymbol\pi_i$ for $i=0,1,...,k-1$}
\label{alg:usim}
\begin{algorithmic}
\Require $n=2$, $N\in\mathbb{N}$ sufficiently large
\Require $X_{1i}$ are independent for $i=0,1,...,k-1$
\Require $X_{2i}=1+r$, $r>-1$ for $i=0,1,...,k-1$
\Require $\boldsymbol\pi_{i}=(q_i(\widehat{W}_i,A_i),1-q_i(\widehat{W}_i,A_i))$ for $i=0,1,...,k-1$
\State $l\gets 0$\Comment{initialize $l$}
\While{$l\leq N$}
\State $l\gets l+1$
\State $i\gets 0$ \Comment{initialize i}
\State $I\gets 1$ \Comment{initialize $I_i$}
\State $A\gets A_0$ \Comment{initialize $A_i$}
\State $\widehat{W}\gets P$ \Comment{initialize $\widehat{W}_i$}
\While{$i\leq k$}
\State $i\gets i+1$
\State $X$ is a realization of $X_{1,i-1}$
\State $Y\gets q_{i-1}(\overline{W},A)\cdot X+(1-q_{i-1}(\overline{W},A))\cdot (1+r)$
\State $B^j$ is a realization of $B_{i-1}^j$ for $j=1,2,...,A$
\State $I\gets I-\sum_{j=1}^IB^j$
\State $A\gets A-\sum_{j=1}^AB^j$
\State $\widehat{W}\gets Y\widehat{W}-IAw_i$ \Comment{computes $\widehat{W}_i$}
\EndWhile
\State $b_l\gets\begin{cases}1,&\widehat{W}\geq 0\\0,&\text{otherwise}\end{cases}$
\EndWhile
\State $\mathbb{P}(\widehat{W}_k\geq w)\gets\frac{1}{N}\sum_{l=1}^Nb_l$\\
\Return{$\mathbb{P}(\widehat{W}_k\geq w)$}
\end{algorithmic}
\end{algorithm}

\begin{algorithm}
\caption{Compute $v_i(x,a)$ and optimal $\boldsymbol\pi_{i}$ for $i=0,1,...,k-1$ and $a=1,2,...,a_0$}
\label{alg:v}
\begin{algorithmic}
\Require $n=2$, $M\in\mathbb{N}$ sufficiently large
\Require $X_{1i}$ are iid and continuous with pdf $f$ and cdf $F$ for $i=0,1,...,k-1$
\Require $X_{2i}=1+r$, $r\geq0$ for $i=0,1,...,k-1$
\Require $G_1=\{.1,.2,...,.9\}$
\State $i\gets k$ \Comment{initialize i}
\While{$i>0$}
\State $i\gets i-1$
\For{$a\in\{1,2,...,a_0\}$}
\State $D_{a,i}\gets\Big\{\frac{jm_{a,i}}{M}:j=1,2,...,M-1\Big\}$
\For{$x\in D_{a,i}$}
\State $q_i^*(x,a)\gets 0$ \Comment{initial proposal for $q_i^*(x,a)$}
\State $v_i(x,a)\gets \eqref{eq:gi0}$\Comment{proposal for $v_i(x,a)$, see \ref{s:valg}}
\State $q_1\gets\underset{q\in G_1}{\arg\max}\ \eqref{eq:maxqh}$\Comment{see \ref{s:valg}}
\State $G_2\gets\{q_1\pm .01j: j=-9,-8,...,10\}$
\State $q_2\gets\underset{q\in G_2}{\arg\max}\ \eqref{eq:maxqh}$
\State $V\gets \eqref{eq:maxqh}\vert_{q=q_2}$
\If{$v_{i}(x,a)<V$}
\State $q_i^*(x,a)\gets q_2$ 
\State $v_i(x,a)\gets V$
\EndIf
\EndFor
\EndFor
\EndWhile\\
\Return{$v_i(x,a),\ \boldsymbol\pi_i=(q_i^*(x,a),1-q_i^*(x,a))$ for $a\in\{1,2,...,a_0\}$, $x\in D_{a,i}$ and $i=0,1,...,k-1$}
\end{algorithmic}
\end{algorithm}

\clearpage
\bibliography{sn-bibliography}


\end{document}